\documentclass{emulateapj}

\usepackage{comment}
\usepackage{ifthen}

\newcommand{\ctbd}[1]{}

\newcommand{\lc}{light curve}
\newcommand{\lcs}{light curves}
\newcommand{\Lc}{Light curve}

\newcommand{\bandd}[1]{\ensuremath{#1}-band}
\newcommand{\band}[1]{\ensuremath{#1} band}

\newcommand{\chisq}{\ensuremath{\chi^2}}

\newcommand{\kms}{\ensuremath{\rm km\,s^{-1}}}
\newcommand{\ms}{\ensuremath{\rm m\,s^{-1}}}

\newcommand{\gcmc}{\ensuremath{\rm g\,cm^{-3}}}
\newcommand{\ergscmsq}{\ensuremath{\rm erg\,s^{-1}\,cm^{-2}}}

\newcommand{\vsini}{\ensuremath{v \sin{i}}}
\newcommand{\feh}{\ensuremath{\rm [Fe/H]}}

\newcommand{\vmac}{\ensuremath{v_{\rm mac}}}
\newcommand{\vmic}{\ensuremath{v_{\rm mic}}}
\newcommand{\rsun}{\ensuremath{R_\sun}}
\newcommand{\msun}{\ensuremath{M_\sun}}
\newcommand{\lsun}{\ensuremath{L_\sun}}

\newcommand{\rstar}{\ensuremath{R_\star}}
\newcommand{\mstar}{\ensuremath{M_\star}}
\newcommand{\lstar}{\ensuremath{L_\star}}

\newcommand{\teffstar}{\ensuremath{T_{\rm eff\star}}}
\newcommand{\rhostar}{\ensuremath{\rho_\star}}
\newcommand{\loggstar}{\ensuremath{\log{g_{\star}}}}

\newcommand{\mearth}{\ensuremath{M_\earth}}

\newcommand{\rpl}{\ensuremath{R_{p}}}
\newcommand{\mpl}{\ensuremath{M_{p}}}

\newcommand{\rhopl}{\ensuremath{\rho_{p}}}

\newcommand{\arstar}{\ensuremath{a/\rstar}}
\newcommand{\zrstar}{\ensuremath{\zeta/\rstar}}

\newcommand{\rjup}{\ensuremath{R_{\rm J}}}
\newcommand{\mjup}{\ensuremath{M_{\rm J}}}

\newcommand{\refsec}[1]{\mbox{\S\ \ref{sec:#1}}}

\newcommand{\reffigl}[1]{Figure~\ref{fig:#1}}
\newcommand{\refsecl}[1]{\mbox{Section \ref{sec:#1}}}

\newcommand{\reftabl}[1]{Table~\ref{tab:#1}}

\newcommand{\flwof}{\mbox{FLWO 1.2\,m}}

\newcommand{\flwos}{\mbox{FLWO 1.5\,m}}

\newcommand{\hatcurCCra}{\ensuremath{17^{\mathrm h}20^{\mathrm m}27.96^{\mathrm s}}}                                  % Right Ascension
\newcommand{\hatcurCCdec}{\ensuremath{+38{\arcdeg}14{\arcmin}31.8{\arcsec}}}                                 % Declination
\newcommand{\hatcurCCtwomass}{2MASS~17202788+3814317}                  % 2MASS identifier
\newcommand{\hatcurCCgsc}{GSC~3086-00152}                              % GSC(1.2) identifier
\newcommand{\hatcurCCtassmv}{9.98}                                    % TASS V-band magnitude
\newcommand{\hatcurLCdip}{\ensuremath{4.5}}                            % BLS detected dip (mmag)
\newcommand{\hatcurLCrprstar}{\ensuremath{0.0805\pm0.0015}}            % Rp/R*
\newcommand{\hatcurLCbsq}{\ensuremath{0.794_{-0.014}^{+0.012}}}        % impact parameter square
\newcommand{\hatcurLCimp}{\ensuremath{0.891_{-0.008}^{+0.007}}}        % impact parameter
\newcommand{\hatcurLCzeta}{\ensuremath{29.53\pm0.42}}                  % zeta/R*
\newcommand{\hatcurLCdur}{\ensuremath{0.0912\pm0.0017}}                % transit duration (days)
\newcommand{\hatcurLCdurhr}{\ensuremath{2.189\pm0.040}}                % transit duration (hours)
\newcommand{\hatcurLCq}{\ensuremath{0.0197\pm0.0004}}                  % fractional transit duration (days)
\newcommand{\hatcurLCingdur}{\ensuremath{0.0287\pm0.0026}}             % ingress/egress duration (days)
\newcommand{\hatcurLCP}{\ensuremath{4.627669\pm0.000005}}              % period (days)
\newcommand{\hatcurLCPshort}{\ensuremath{4.6277}}                      % period (days)
\newcommand{\hatcurLCT}{\ensuremath{2,\!454,\!875.28938\pm0.00047}}          % epoch (BJD)
\newcommand{\hatcurLCTA}{\ensuremath{2,\!452,\!728.0512\pm0.0024}}           % TA (BJD)
\newcommand{\hatcurLCTB}{\ensuremath{2,\!454,\!967.8427\pm0.0005}}           % TB (BJD)
\newcommand{\hatcurLCiblendnew}{\ensuremath{0.73\pm0.09}}                 % HATNet iblend factor
\newcommand{\hatcurLCiblendold}{\ensuremath{0.96\pm0.04}}                 % HATNet iblend factor
\newcommand{\hatcurSMEiteff}{\ensuremath{6310\pm80}}                   % Ini SME, stellar effective temperature
\newcommand{\hatcurSMEizfeh}{\ensuremath{-0.02\pm0.05}}                % Ini SME, stellar metallicity
\newcommand{\hatcurSMEizfehshort}{\ensuremath{-0.02}}                  % Ini SME, stellar metallicity
\newcommand{\hatcurSMEilogg}{\ensuremath{3.80\pm0.10}}                  % Ini SME, stellar surface gravity
\newcommand{\hatcurSMEivsin}{\ensuremath{8.8\pm0.5}}                   % Ini SME, stellar rotational velocity
\newcommand{\hatcurSMEivmac}{\ensuremath{0.0}}                         % Ini SME, stellar macroturbulence
\newcommand{\hatcurSMEivmic}{\ensuremath{0.85}}                        % Ini SME, stellar microturbulence
\newcommand{\hatcurSMEiiteff}{\ensuremath{6600\pm90}}                  % Final SME, stellar effective temperature
\newcommand{\hatcurSMEiizfeh}{\ensuremath{+0.11\pm0.08}}               % Final SME, stellar metallicity
\newcommand{\hatcurSMEiizfehshort}{\ensuremath{+0.11}}                 % Final SME, stellar metallicity
\newcommand{\hatcurSMEiilogg}{\ensuremath{4.21\pm0.00}}                % Final SME, stellar surface gravity
\newcommand{\hatcurSMEiivsin}{\ensuremath{8.4\pm0.5}}                  % Final SME, stellar rotational velocity
\newcommand{\hatcurSMEiivmac}{\ensuremath{5.30}}                       % Final SME, stellar macroturbulence
\newcommand{\hatcurSMEiivmic}{\ensuremath{0.85}}                       % Final SME, stellar microturbulence
\newcommand{\hatcurDSteff}{\ensuremath{6500\pm100}}                    % DS stellar effective temperature
\newcommand{\hatcurDSlogg}{\ensuremath{3.5\pm0.25}}                    % DS stellar surface gravity
\newcommand{\hatcurDSvsini}{\ensuremath{11.7\pm1.0}}                   % DS stellar rotational velocity
\newcommand{\hatcurDSgamma}{\ensuremath{-20.81\pm0.28}}                % DS absolute gamma velocity
\newcommand{\hatcurDSrvrms}{\ensuremath{0.28}}                         % DS rms of RV values [km/s]
\newcommand{\hatcurLBii}{\ensuremath{0.1089}}                          % Limb darkening parameters, Gamma1, i-band
\newcommand{\hatcurLBiii}{\ensuremath{0.2439}}                         % Limb darkening parameters, Gamma2, i-band
\newcommand{\hatcurISOmlong}{\ensuremath{1.386\pm0.045}}               % stellar mass
\newcommand{\hatcurISOrlong}{\ensuremath{1.468\pm0.054}}               % stellar radius
\newcommand{\hatcurISOlogg}{\ensuremath{4.25\pm0.03}}                  % stellar surface gravity from isochrones
\newcommand{\hatcurISOlum}{\ensuremath{3.66\pm0.37}}                   % stellar luminosity
\newcommand{\hatcurISOmv}{\ensuremath{3.31\pm0.12}}                    % stellar absolute magnitude
\newcommand{\hatcurISOage}{\ensuremath{1.3\pm0.4}}                     % stellar age
\newcommand{\hatcurISOageshort}{\ensuremath{1.3}}                     % stellar age
\newcommand{\hatcurISOspec}{F}                                         % stellar spectral type
\newcommand{\hatcurRVK}{\ensuremath{219.0\pm3.3}}                      % RV semi-amplitude [m/s]
\newcommand{\hatcurRVk}{\ensuremath{-0.009\pm0.009}}                   % e*cos(omega)
\newcommand{\hatcurRVh}{\ensuremath{0.106\pm0.013}}                    % e*sin(omega)
\newcommand{\hatcurRVgamma}{\ensuremath{26.6\pm2.4}}                   % RV gamma velocity, relative scale [m/s]
\newcommand{\hatcurRVjitter}{\ensuremath{7.3}}                         % RV jitter (m/s)
\newcommand{\hatcurRVeccen}{\ensuremath{0.107\pm0.013}}                % eccentricity
\newcommand{\hatcurRVomega}{\ensuremath{94\pm4}}                       % argument of pericenter
\newcommand{\hatcurPPi}{\ensuremath{83.5\pm0.3}}                       % orbital inclination
\newcommand{\hatcurPPlogg}{\ensuremath{3.621\pm0.037}}                   % planetary surface gravity (log cgs)
\newcommand{\hatcurPPar}{\ensuremath{8.87\pm0.29}}                     % relative orbital radius (a/R*)
\newcommand{\hatcurPParel}{\ensuremath{0.0606\pm0.0007}}               % semimajor axis (AU)
\newcommand{\hatcurPPrho}{\ensuremath{1.82\pm0.24}}                    % planetary density (cgs)
\newcommand{\hatcurPPmshort}{\ensuremath{2.23}}                        % planetary mass (M_jup)
\newcommand{\hatcurPPmlong}{\ensuremath{2.232\pm0.059}}                % planetary mass (M_jup)
\newcommand{\hatcurPPrlong}{\ensuremath{1.150\pm0.052}}                % planetary radius (R_jup)
\newcommand{\hatcurPPmrcorr}{\ensuremath{0.45}}                        % mass/radius correlation
\newcommand{\hatcurPPteff}{\ensuremath{1570\pm34}}                     % planetary temperature (K)
\newcommand{\hatcurPPtheta}{\ensuremath{0.168\pm0.008}}                % Safranov number
\newcommand{\hatcurPPfluxavg}{\ensuremath{1.37 \pm 0.12}}        % flux on average (CGS)
\newcommand{\hatcurXdist}{\ensuremath{205\pm11}}                        % distance (pc)
\newcommand{\hatcur}{HAT-P-14}
\newcommand{\hatcurb}{HAT-P-14b}

\newcommand{\hatcurSMEversion}{ii}                                       % Final SME version:i or ii?
\newcommand{\hatcurSMEteff}{\ifthenelse{\equal{\hatcurSMEversion}{i}}{\hatcurSMEiteff}{\hatcurSMEiiteff}}
\newcommand{\hatcurSMEzfeh}{\ifthenelse{\equal{\hatcurSMEversion}{i}}{\hatcurSMEizfeh}{\hatcurSMEiizfeh}}
\newcommand{\hatcurSMEzfehshort}{\ifthenelse{\equal{\hatcurSMEversion}{i}}{\hatcurSMEizfehshort}{\hatcurSMEiizfehshort}}
\newcommand{\hatcurSMElogg}{\ifthenelse{\equal{\hatcurSMEversion}{i}}{\hatcurSMEilogg}{\hatcurSMEiilogg}}
\newcommand{\hatcurSMEvsin}{\ifthenelse{\equal{\hatcurSMEversion}{i}}{\hatcurSMEivsin}{\hatcurSMEiivsin}}
\newcommand{\hatcurSMEvmac}{\ifthenelse{\equal{\hatcurSMEversion}{i}}{\hatcurSMEivmac}{\hatcurSMEiivmac}}
\newcommand{\hatcurSMEvmic}{\ifthenelse{\equal{\hatcurSMEversion}{i}}{\hatcurSMEivmic}{\hatcurSMEiivmic}}

\newcommand{\hatcurisoshort}{YY}
\newcommand{\hatcurlumind}{\arstar}
\newcommand{\titledag}{$\dagger$}

\shortauthors{Torres et al.}
\shorttitle{\hatcur\lowercase{b}}

\begin{document}

%% Titlepage
\title{\hatcur\lowercase{b}: A 2.2\,\mjup\ exoplanet transiting a bright
\hatcurISOspec\ star \altaffilmark{\titledag}}

\author{
	G.\ Torres\altaffilmark{1},
	G.\ \'A.\ Bakos\altaffilmark{1,2},
	J.~Hartman\altaffilmark{1},
	G\'eza~Kov\'acs\altaffilmark{3},
	R.\ W.\ Noyes\altaffilmark{1},
	D.\ W.\ Latham\altaffilmark{1},
	D.\ A.\ Fischer\altaffilmark{4},
	J.\ A.\ Johnson\altaffilmark{5},
	G.\ W.\ Marcy\altaffilmark{6},
	A.\ W.\ Howard\altaffilmark{6},
	D.\ D.\ Sasselov\altaffilmark{1},
        D.\ Kipping\altaffilmark{1,7},
	B.\ Sip\H{o}cz\altaffilmark{1,8},
	R.\ P.\ Stefanik\altaffilmark{1},
	G.\ A.\ Esquerdo\altaffilmark{1},
        M.\ E.\ Everett\altaffilmark{9},
	J.\ L\'az\'ar\altaffilmark{10},
	I.\ Papp\altaffilmark{10},
	P.\ S\'ari\altaffilmark{10}
}

\altaffiltext{1}{Harvard-Smithsonian Center for Astrophysics,
	Cambridge, MA; e-mail: gtorres@cfa.harvard.edu}

\altaffiltext{2}{NSF Fellow}

\altaffiltext{3}{Konkoly Observatory, Budapest, Hungary}

\altaffiltext{4}{Department of Physics and Astronomy, San Francisco
	State University, San Francisco, CA}

\altaffiltext{5}{Institute for Astronomy, University of Hawaii,
Honolulu, HI; NSF Postdoctoral Fellow}

\altaffiltext{6}{Department of Astronomy, University of California,
	Berkeley, CA}

\altaffiltext{7}{Department of Physics and Astronomy, University
College London, UK}

\altaffiltext{8}{Centre for Astrophysics Research, University of
Hertfordshire, Hatfield, UK}

\altaffiltext{9}{Planetary Science Institute, Tucson, AZ}

\altaffiltext{10}{Hungarian Astronomical Association, Budapest, 
	Hungary}

\altaffiltext{$\dagger$}{
	Based in part on observations obtained at the W.\ M.\ Keck
	Observatory, which is operated by the University of California and
	the California Institute of Technology.
}

\begin{abstract}

We report the discovery of \hatcurb{}, a fairly massive transiting
extrasolar planet orbiting the moderately bright star \hatcurCCgsc\
($V = \hatcurCCtassmv$), with a period of $P=\hatcurLCP$\,d. The
transit is close to grazing (impact parameter \hatcurLCimp) and has a
duration of \hatcurLCdur\,d, with a reference epoch of mid transit of
$T_c = \hatcurLCT$ (BJD). The orbit is slightly eccentric ($e =
\hatcurRVeccen$), and the orientation is such that occultations are
unlikely to occur.  The host star is a slightly evolved
mid-\hatcurISOspec\ dwarf with a mass of \hatcurISOmlong\,\msun, a
radius of \hatcurISOrlong\,\rsun, effective temperature
\hatcurSMEteff\,K, and a slightly metal-rich composition corresponding
to $\feh = \hatcurSMEzfeh$. The planet has a mass of
\hatcurPPmlong\,\mjup\ and a radius of \hatcurPPrlong\,\rjup, implying
a mean density of \hatcurPPrho\,\gcmc. Its radius is well reproduced
by theoretical models for the \hatcurISOageshort\,Gyr age of the
system if the planet has a heavy-element fraction of about
50\,\mearth\ (7\% of its total mass).  The brightness, near-grazing
orientation, and other properties of \hatcur{} make it a favorable
transiting system to look for changes in the orbital elements or
transit timing variations induced by a possible second planet, and
also to place meaningful constraints on the presence of sub-Earth mass
or Earth mass exomoons, by monitoring it for transit duration
variations.

\end{abstract}


\section{Introduction}
\label{sec:introduction}

More than five dozen transiting extrasolar planets (TEPs) have been
discovered to date, by both ground-based surveys and, recently,
space-based surveys such as the {\em CoRoT} and {\em Kepler} missions
\citep{baglin:2006, borucki:2010}. The ground-based surveys are
typically able to detect TEPs orbiting only the brighter stars.
However this disadvantage has the compensatory advantage that those
TEPs detected are amenable to a wide range of follow-up studies, which
can provide valuable insight into their atmospheric properties and
other physical conditions. 

Among the ground-based surveys, the Hungarian-made Automated Telescope
Network \citep[HATNet;][]{bakos:2004} survey has been one of the main
contributors to the discovery of TEPs. In operation since 2003, it has
now covered approximately 11\% of the sky, searching for TEPs around
bright stars ($8\lesssim I \lesssim 12.5$). HATNet operates six
wide-field instruments: four at the Fred Lawrence Whipple Observatory
(FLWO) in Arizona, and two on the roof of the hangar servicing the
Smithsonian Astrophysical Observatory's Submillimeter Array, in
Hawaii.  Since 2006, HATNet has announced and published 13 TEPs. In
this work we report our fourteenth discovery, around the relatively
bright ($V = \hatcurCCtassmv$) star we refer to as \hatcur, also known
as \hatcurCCgsc{}.

The layout of the paper is as follows. In \refsecl{obs} we report the
detection of the photometric signal and the follow-up spectroscopic
and photometric observations of \hatcur{}. In \refsecl{bisec} we
examine the spectroscopic evidence to confirm the planetary nature of
the object.  \refsecl{analysis} describes the analysis of the data,
beginning with the determination of the stellar parameters, and
continuing with a description of our global modeling of the photometry
and radial velocities. Results are presented in \refsecl{results}, and
discussed in \refsecl{discussion}.

\section{Observations}
\label{sec:obs}

\subsection{Photometric detection}
\label{sec:detection}

The transits of \hatcurb{} were detected with the HAT-10 telescope
located in Arizona. Observations in a field containing the star
\hatcurCCgsc{} were made on a nightly basis between 2005 May and July,
as permitted by weather conditions. We gathered 1246 exposures of
150\,s with a cadence of 3.2 minutes, 206 exposures of 180\,s at 3.7
minute cadence, and 1991 exposures of 300\,s at a 5.7 minute cadence.
The reason for the somewhat inhomogeneous data set is that the
observations were acquired during the commissioning phase of HAT-10.
The first exposures with 150\,s cadence were taken with an Apogee
ALTA~E42 back-illuminated CCD through a Cousins \bandd{I}
filter. Exposure times were increased to 180\,s after some initial
testing. Finally, the CCD was replaced with an Apogee E10
front-illuminated model, and exposure times were further increased to
300 seconds.  Each image contained approximately 22,000 stars down to
a brightness limit of $I \sim 14.0$. For the brightest stars in the
field we achieved a photometric precision of 3.5\,mmag per image in
the combined \lc.

\begin{figure}[!t]
\epsscale{1.12}
\plotone{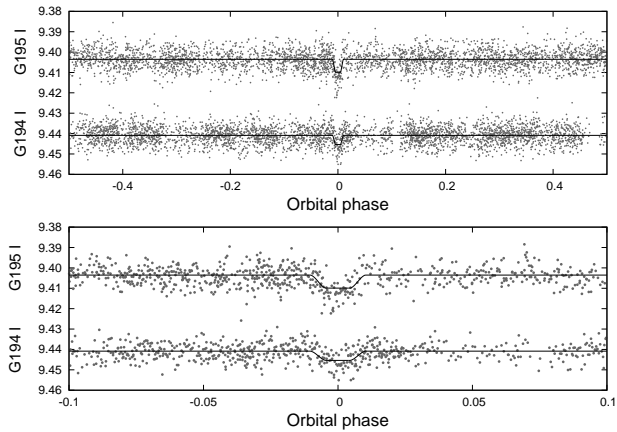}
\caption{ {\it Top:} Unbinned \bandd{I} \lcs{} of \hatcur{} obtained
  with the HAT-5 and HAT-10 nodes of HATNet, showing all 3189 and 3443
  instrumental measurements from fields G195 and G194, respectively.
  The photometry is phase-folded with the period $P = 4.627669$ days
  resulting from the global fit described in \refsecl{analysis}. The
  solid lines show the transit model fit to the light curves
  (\refsecl{globmod}). {\it Bottom:} The same, for phases within 0.1
  of the transit center.
\label{fig:hatnet}}
\end{figure}

The calibration of the HATNet frames was carried out using standard
photometric procedures. The calibrated images were then subjected to
star detection and an astrometric solution, as described in more
detail by \cite{pal:2006}. Aperture photometry was performed on each
image at the stellar centroids derived from the Two Micron All Sky
Survey catalog \citep[2MASS;][]{skrutskie:2006} and the individual
astrometric solutions. The resulting \lcs\ were decorrelated (cleaned
of trends) using the External Parameter Decorrelation technique in
``constant'' mode \citep[EPD; see][]{bakos:2009} and the Trend
Filtering Algorithm \citep[TFA; see][]{kovacs:2005}. The \lcs{} were
then searched for periodic box-shaped signals using the Box
Least-Squares method \citep[BLS;][]{kovacs:2002}. We detected a
significant signal in the \lc{} of \hatcurCCgsc{} (also known as
\hatcurCCtwomass{}; $\alpha = \hatcurCCra$, $\delta = \hatcurCCdec$;
J2000; $V=\hatcurCCtassmv$; \citealp{droege:2006}), with an apparent
depth of \hatcurLCdip\,mmag, and a period of $P=\hatcurLCPshort$\,days
(see \reffigl{hatnet}).  The drop in brightness had a
first-to-last-contact duration $q$, relative the total period $P$, of
$q = \hatcurLCq$, corresponding to a total duration of $Pq =
\hatcurLCdurhr$~hr.

\hatcur\ falls in an overlapping area between two HATNet fields
internally labeled G194 and G195. While the discovery was made on the
basis of the G194 \lc, the signal was later recovered in the 3189
\bandd{I} observations from field G195. These latter data were taken
during the commissioning phase of the HAT-5 telescope at FLWO, in
early 2003, making this one of the TEPs with the longest photometric
coverage. Curiously, the signal is present in the early HAT-5 (G195)
data with higher significance than in field G194. The discovery of
\hatcur\ in 2003 was probably missed due to a combination of factors,
including our different criteria for scheduling follow-up observations
at the time, and the lack of the Trend Filtering Algorithm, which was
only implemented and brought into regular use in 2005.

\subsection{Reconnaissance spectroscopy}
\label{sec:recspec}

As is routine in the HATNet project, all candidates are subjected to
careful scrutiny before investing valuable time on large telescopes.
This includes spectroscopic observations at relatively modest
facilities to establish whether the transit-like feature in the light
curve of a candidate might be due to astrophysical phenomena other
than a planet transiting a star. Many of these false positives are
associated with large radial-velocity variations in the star (tens of
\kms) that are easily recognized. For this we have made use here of
the Harvard-Smithsonian Center for Astrophysics (CfA) Digital
Speedometer \citep[DS;][]{latham:1992}, an echelle spectrograph
mounted on the \flwos\ telescope. This instrument delivers
high-resolution spectra ($\lambda/\Delta\lambda \approx 35,\!000$)
over a single order centered on the \ion{Mg}{1}\,b triplet
($\sim$5187\,\AA), with typically low signal-to-noise (S/N) ratios
that are nevertheless sufficient to derive radial velocities (RVs)
with moderate precisions of 0.5--1.0\,\kms\ for slowly rotating stars.
The same spectra can be used to estimate the effective temperature,
surface gravity, and projected rotational velocity of the host star,
as described by \cite{torres:2002}.  With this facility we are able to
reject many types of false positives, such as F dwarfs orbited by M
dwarfs, grazing eclipsing binaries, or triple or quadruple star
systems. Additional tests are performed with other spectroscopic
material described in the next section.

For \hatcur{} we obtained four
observations with the DS between April and May of 2008.  The velocity
measurements showed an rms residual of \hatcurDSrvrms\,\kms,
consistent with no detectable RV variation within the precision of the
measurements. All spectra were single-lined, i.e., there is no
evidence for additional stars in the system.  The atmospheric
parameters we infer from these observations are the following:
effective temperature $\teffstar = \hatcurDSteff\,K$, surface gravity
$\loggstar = \hatcurDSlogg$ (log cgs), and projected rotational
velocity $\vsini = \hatcurDSvsini\,\kms$. The effective temperature
corresponds to a mid-\hatcurISOspec\ dwarf. The mean heliocentric RV
of \hatcur\ is $\gamma_{\rm RV} = \hatcurDSgamma$\,\kms.

\subsection{High resolution, high S/N spectroscopy}
\label{sec:hispec}

Given the statistically significant transit detection by HATNet, and
the encouraging DS results that rule out obvious false positives, we
proceeded with the follow-up of this candidate by obtaining
high-resolution, high-S/N spectra in order to characterize the RV
variations with higher precision, and to refine the determination of
the stellar parameters.  For this we used the HIRES instrument
\citep{vogt:1994} on the Keck~I telescope located on Mauna Kea,
Hawaii, between 2008 May 15 and 2009 October 1. The width of the
spectrometer slit was $0\farcs86$, resulting in a resolving power of
$\lambda/\Delta\lambda \approx 55,\!000$, with a wavelength coverage
of $\sim$3800--8000\,\AA\@.

\begin{figure} [ht]
\epsscale{1.12}
\plotone{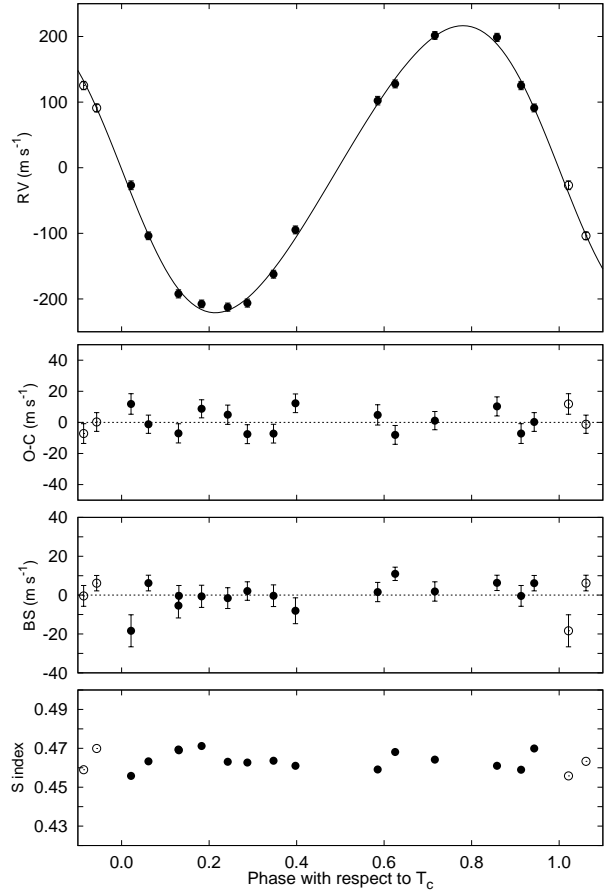}
\caption{
    {\em Top panel}: Keck/HIRES RV measurements for \hbox{\hatcur{}}
    shown as a function of orbital phase, along with our best-fit
    model (see \reftabl{planetparam}). Zero phase corresponds to the
    time of mid-transit. The center-of-mass velocity has been
    subtracted.
	{\em Second panel}: Velocity $O\!-\!C$ residuals from the best
        fit.  The error bars (which are too small to be seen in the
        top panel) include a component from astrophysical jitter
        ($\hatcurRVjitter$\,\ms) added in quadrature to the formal
        errors (see \refsecl{results}).
	{\em Third panel}: Bisector spans (BS), with the mean value
    subtracted. The measurement from the template spectrum is
    included.
	{\em Bottom panel}: Relative chromospheric activity index $S$
	measured from the Keck spectra.
    Note the different vertical scales of the panels. Observations shown
    twice are represented with open symbols.
\label{fig:rvbis}}
\end{figure}

We obtained a total of 14 exposures through an iodine gas absorption
cell, which was used to superimpose a dense forest of $\mathrm{I}_2$
lines on the stellar spectrum and establish an accurate wavelength
fiducial \citep[see][]{marcy:1992}. An additional exposure was taken
without the iodine cell, for use as a template in the reductions.
Relative RVs in the solar system barycentric frame were derived as
described by \cite{butler:1996}, incorporating full modeling of the
spatial and temporal variations of the instrumental profile. The RV
measurements and their uncertainties are listed in \reftabl{rvs}. The
period-folded data, along with our best fit described below in
\refsecl{analysis}, are displayed in \reffigl{rvbis}.

\begin{deluxetable}{lrcrcc}
\tablewidth{0pc}
\tablecaption{
	Relative radial velocities, bisector spans, and activity index
	measurements of \hatcur{}.
	\label{tab:rvs}
}
\tablehead{
	\colhead{BJD} & 
	\colhead{RV\tablenotemark{a}} & 
	\colhead{\ensuremath{\sigma_{\rm RV}}\tablenotemark{b}} & 
	\colhead{BS} & 
	\colhead{\ensuremath{\sigma_{\rm BS}}} & 
	\colhead{$S$\tablenotemark{c}} \\ 
	\colhead{\hbox{(2,454,000$+$)}} & 
	\colhead{(\ms)} & 
	\colhead{(\ms)} &
	\colhead{(\ms)} &
    \colhead{(\ms)} &
	\colhead{} 
}
\startdata
 $ \phn602.85803 $          & $  -193.02 $ & $     3.63 $ & $    -5.39 $ & $     6.36 $ & $    0.469  $  \\
 $ \phn602.86212 $ \tablenotemark{d}  & \nodata  & \nodata  & $   -0.39 $ & $     5.33 $ & $    0.469  $  \\
 $ \phn603.10266 $          & $  -208.31 $ & $     2.97 $ & $    -0.62 $ & $     5.69 $ & $    0.471  $  \\
 $ \phn603.86301 $          & $  -163.07 $ & $     3.39 $ & $    -0.31 $ & $     5.59 $ & $    0.464  $  \\
 $ \phn604.09553 $          & $   -95.63 $ & $     3.35 $ & $    -8.06 $ & $     6.65 $ & $    0.461  $  \\
 $ \phn633.99341 $          & $   197.89 $ & $     3.59 $ & $     6.31 $ & $     3.95 $ & $    0.461  $  \\
 $ \phn634.93450 $          & $  -104.57 $ & $     3.08 $ & $     6.21 $ & $     4.03 $ & $    0.463  $  \\
 $ \phn635.98053 $          & $  -207.08 $ & $     3.48 $ & $     2.08 $ & $     4.76 $ & $    0.463  $  \\
 $ \phn637.96376 $          & $   200.73 $ & $     3.10 $ & $     1.88 $ & $     4.99 $ & $    0.464  $  \\
 $ \phn639.01426 $          & $    90.31 $ & $     3.34 $ & $     6.15 $ & $     3.97 $ & $    0.470  $  \\
 $ \phn641.98653 $          & $   101.38 $ & $     4.20 $ & $     1.57 $ & $     4.96 $ & $    0.459  $  \\
 $ \phn777.70610 $          & $   124.49 $ & $     3.94 $ & $    -0.40 $ & $     5.36 $ & $    0.459  $  \\
 $ \phn955.08247 $          & $  -213.27 $ & $     3.65 $ & $    -1.55 $ & $     5.37 $ & $    0.463  $  \\
 $ 1017.01279 $             & $   126.95 $ & $     3.44 $ & $    10.95 $ & $     3.44 $ & $    0.468  $  \\
 $ 1106.77226 $             & $   -27.59 $ & $     4.34 $ & $   -18.36 $ & $     8.23 $ & $    0.456  $  \\ [-1.5ex]
\enddata
\tablenotetext{a}{
	The zero-point of these velocities is arbitrary. An
	overall offset $\gamma_{\rm rel}$ fitted to these velocities in
	\refsecl{globmod} has {\em not} been subtracted.
}
\tablenotetext{b}{
        Internal errors excluding the component of astrophysical jitter
	considered in \refsecl{results}.
}
\tablenotetext{c}{
        Relative chromospheric activity index, not calibrated to the
        scale of \citet{vaughan:1978}.
}
\tablenotetext{d}{This observation corresponds to the iodine-free
template, which does not yield a RV measurement but does permit the BS
and $S$ index to be computed.}

\end{deluxetable}

In the same figure we show also the relative $S$ index, which is a
measure of the chromospheric activity of the star derived from the
flux in the cores of the \ion{Ca}{2} H and K lines.  This index was
computed following the prescription given by \citet{vaughan:1978},
after matching each spectrum to a reference spectrum using a
transformation that includes a wavelength shift and a flux scaling
that is a polynomial as a function of wavelength. The transformation
was determined on regions of the spectra that are not used in
computing this indicator. Note that our relative $S$ index has not
been calibrated to the scale of \citet{vaughan:1978}. We do not detect
any significant variation of the index correlated with orbital phase;
such a correlation might have indicated that the RV variations could
be due to stellar activity, casting doubt on the planetary nature of
the candidate. There is no sign of emission in the cores of the
\ion{Ca}{2} H and K lines in any of our spectra, from which we
conclude that the chromospheric activity level in \hatcur{} is very
low.

\subsection{Photometric follow-up observations}
\label{sec:phot}

\begin{figure}[!ht]
\epsscale{1.12}
\plotone{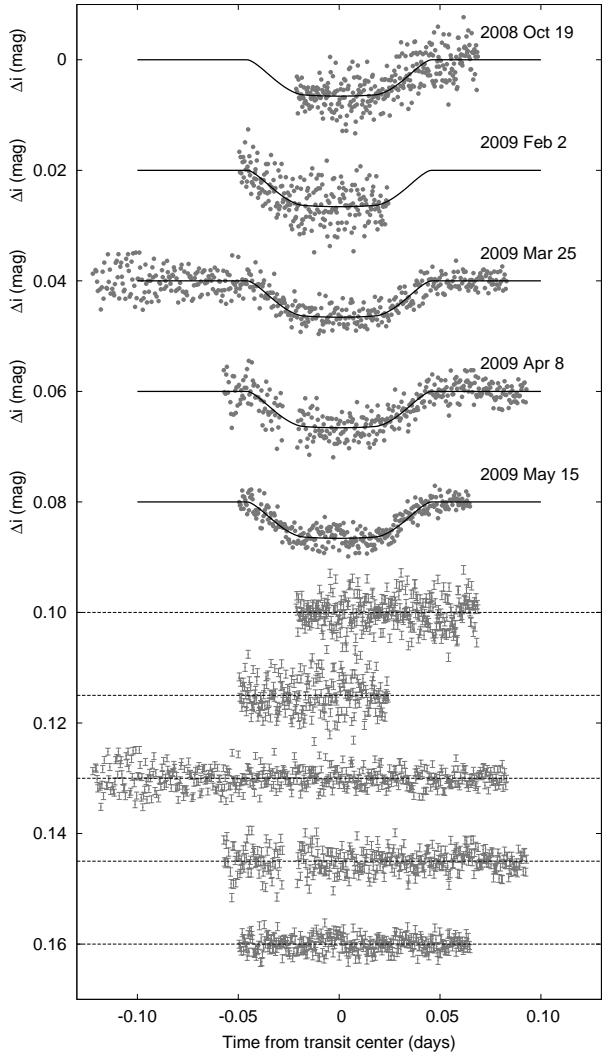}
\caption{
	Unbinned instrumental transit \lcs\ in the Sloan \band{i},
        acquired with KeplerCam on the \flwof{} telescope.  The light
        curves have been processed with EPD and TFA, as described in
        \refsec{globmod}.  The dates of the events are indicated.
        Curves after the first are displaced vertically for clarity.
        Our best fit from the global modeling described in
        \refsecl{globmod} is shown by the solid lines.  Residuals from
        the fits are displayed at the bottom, in the same order as the
        top curves.  The error bars represent the photon and
        background shot noise, plus the readout noise.
\label{fig:lc}}
\end{figure}

In order to permit a more accurate modeling of the light curve, we
conducted additional photometric observations with the KeplerCam CCD
camera on the \flwof{} telescope. We observed five transit events of
\hatcur{} on the nights of 2008 October 19, 2009 February 2, 2009
March 25, 2009 April 8, and 2009 May 15 (\reffigl{lc}). A Sloan
\bandd{i} filter was used for all observations. The number of images
we acquired on each of these events is 326, 239, 470, 359, and 304,
respectively. The exposure times were 10, 13, 20, 15, and 15\,s,
resulting in a cadence of 22, 25, 32, 27, and 27\,s.  The reduction of
these images, including basic calibration, astrometry, aperture
photometry, and trend removal, was performed as described by
\cite{bakos:2009}.  The final time series are shown in the top portion
of \reffigl{lc}, along with our best-fit transit \lc{} model described
below; the individual measurements are reported in \reftabl{phfu}.

\begin{deluxetable}{lccc}
\tablewidth{0pc}
\tablecaption{High-precision differential photometry of \hatcur\label{tab:phfu}}
\tablehead{
	\colhead{BJD} & 
	\colhead{$\Delta i$ \tablenotemark{a}} & 
	\colhead{\ensuremath{\sigma_{\Delta i}}} &
	\colhead{Orig. Mag \tablenotemark{b}} \\
	\colhead{\hbox{~~~~(2,454,000$+$)~~~~}} & 
	\colhead{(mag)} & 
	\colhead{(mag)} &
	\colhead{(mag)} 
}
\startdata
$ 759.57643 $\dotfill & $   0.00505 $ & $   0.00076 $ & $   8.54130 $ \\
$ 759.57670 $\dotfill & $   0.00281 $ & $   0.00076 $ & $   8.53918 $ \\
$ 759.57697 $\dotfill & $   0.00561 $ & $   0.00075 $ & $   8.54200 $ \\
$ 759.57725 $\dotfill & $   0.00948 $ & $   0.00076 $ & $   8.54627 $ \\
$ 759.57754 $\dotfill & $   0.00333 $ & $   0.00075 $ & $   8.53897 $ \\ [-1.5ex]
\enddata
\tablenotetext{a}{
	The out-of-transit level has been subtracted. These magnitudes have
	been subjected to the EPD and TFA procedures, carried out
	simultaneously with the transit fit.
}
\tablenotetext{b}{
	Raw magnitude values without application of the EPD
	and TFA procedures.
}
\tablecomments{
    This table is available in a machine-readable form in the
    online journal. A portion is shown here for guidance regarding
    its form and content.
}
\end{deluxetable}

\section{False positive rejection}
\label{sec:bisec}

Our initial spectroscopic analyses discussed in \refsecl{recspec} and
\refsecl{hispec} rule out the most obvious astrophysical false
positive scenarios. However, more subtle phenomena such as blends
(contamination by an unresolved eclipsing binary, whether in the
background or associated with the target) can still mimic both the
photometric and spectroscopic signatures we see.  Following
\cite{torres:2007}, we explore here the possibility that the measured
Keck/HIRES radial velocities are not real, but are instead caused by
distortions in the spectral line profiles due to this contamination.
We determined the spectral line bisectors and corresponding bisector
spans (a measure of the line asymmetry) from each of our Keck spectra
as described in \S 5 of \cite{bakos:2007}. The bisector spans are
displayed in the third panel of \reffigl{rvbis}. At first glance there
is little variation as a function of orbital phase, in contrast to the
$\sim$400\,\ms\ peak-to-peak change in the RVs shown in the top
panel. However, closer examination gives the visual impression of a
slight correlation with the radial velocities, in the sense that the
bisector spans for the first half of the orbit tend to be zero or
negative (as are the RVs), whereas those in the second half tend to be
zero or positive (again following the RVs).

\begin{figure}[!hb]
\epsscale{1.12}
\plotone{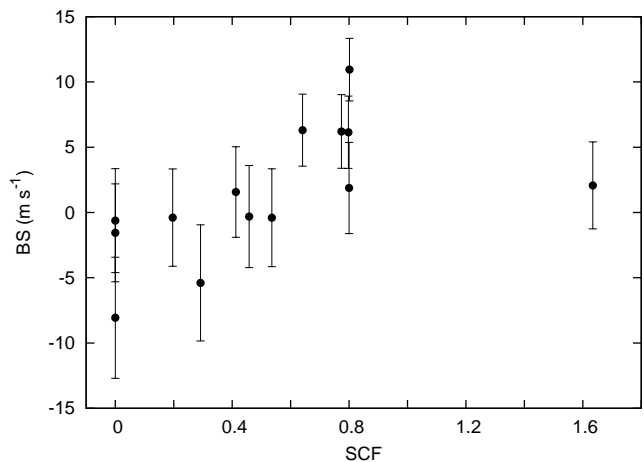}

\caption{
Bisector spans (BS) shown as a function of the sky contamination
factor \citep[SCF; see][]{hartman:2009}, which is a measure of the
influence of scattered moonlight in each of our spectra (see text).
One outlier has been removed, corresponding to the last Keck/HIRES
observation listed in \reftabl{rvs}, which was obtained under cloudy
conditions. The correlation is statistically significant, as indicated
by a Spearman rank-order correlation false-alarm probability of
0.04\%.
\label{fig:scf}}
\end{figure}

A Spearman rank-order correlation test indicates that a correlation as
strong as that suggested between the bisector spans and the RVs would
occur by chance 9.5\% of the time (false-alarm probability, FAP),
which we do not consider to be small enough to confirm the apparent
correlation at a significant level. If we remove one outlier among the
bisector spans near phase 0.0, corresponding to a spectrum taken
through considerable clouds, the FAP is reduced to 6.4\%, still not
very compelling.  We examined this further by considering the
possibility that the slight apparent changes in the line profiles
might be caused by contamination not from an unresolved eclipsing
binary, but from scattered moonlight under partly cloudy observing
conditions. Evidence of this effect was seen in one of our previous
discoveries, HAT-P-12 \citep[][]{hartman:2009}.  Following the same
procedures used there, we computed the sky contamination factor (SCF)
for each of our spectra of \hatcur{}, which quantifies the sky
brightness due to moonlight and also accounts for the velocity
difference between the moon and the star. We then contrasted the SCF
values against the corresponding bisector spans.  This is shown in
\reffigl{scf}. A rather clear trend seems to emerge, suggesting that
the distortions in the line shapes as measured by the bisector spans
could indeed be due to contamination from moonlight. A Spearman
rank-order correlation test excluding the same outlier as before
confirms this, giving a FAP of only 0.04\%. The apparent correlation
between the bisector spans and the velocities in \reffigl{rvbis}
reflects the fact that the six spectra taken during the second half of
the orbit happen to have SCF values that are larger, on average, than
those in the first half.  Based on the above, we conclude that the
small changes in the line shapes as measured by the highly sensitive
bisector spans are well explained by contamination from scattered
moonlight, and that there is no real correlation with the radial
velocities. The radial velocity variations in \hatcur{} are therefore
real, and must be due to a planetary companion in orbit around the
star.

\section{Analysis}
\label{sec:analysis}

\subsection{Properties of the parent star}
\label{sec:stelparam}

Fundamental parameters of the host star \hatcur{} such as the mass
(\mstar) and radius (\rstar), which are needed to infer the planetary
properties, depend strongly on other stellar quantities that can be
derived spectroscopically.  For this we have relied on our template
spectrum obtained with the Keck/HIRES instrument, and the analysis
package known as Spectroscopy Made Easy \citep[SME;][]{valenti:1996},
along with the atomic line database of \cite{valenti:2005}. This
analysis yielded the following {\em initial} values:
effective temperature $\teffstar=\hatcurSMEiteff$\,K, 
surface gravity $\loggstar=\hatcurSMEilogg$\,(cgs),
metallicity $\feh=\hatcurSMEizfeh$\,dex, and 
projected rotational velocity $\vsini=\hatcurSMEivsin\,\kms$.
The temperature and metallicity uncertainties have been conservatively
increased by a factor of two over their nominal values, based on
previous experience, to account for possible systematic errors.

In principle the effective temperature and metallicity, along with the
surface gravity taken as a luminosity indicator, could be used as
constraints to infer the stellar mass and radius by comparison with
stellar evolution models.  However, the effect of \loggstar\ on the
spectral line shapes is rather subtle, and as a result it is typically
difficult to determine accurately, so that it is a rather poor
luminosity indicator for this particular application. A trigonometric
parallax is unfortunately unavailable for \hatcur{} since the star was
not included among the targets of the {\it Hipparcos\/} mission
\citep{perryman:1997}, despite being bright enough. In any case, for
planetary transits a stronger constraint on the luminosity is often
provided by \arstar, the normalized semimajor axis, which is closely
related to \rhostar, the mean stellar density. The quantity \arstar\
can be derived directly from the transit \lcs\ \citep[see][and also
\refsecl{globmod}]{sozzetti:2007}. This, in turn, allows us to improve
on the determination of the spectroscopic parameters by supplying an
indirect constraint on the weakly determined spectroscopic value of
\loggstar\ that removes degeneracies. We take this approach here, as
described below.

\begin{deluxetable}{lcl}[!h]
\tablewidth{0pc}
\tabletypesize{\scriptsize}
\tablecaption{
	Stellar parameters for \hatcur{}
	\label{tab:stellar}
}
\tablehead{
	\colhead{~~~~~~~~Parameter~~~~~~~~}	&
	\colhead{Value} &
	\colhead{Source}
}
\startdata
\noalign{\vskip -3pt}
\sidehead{Spectroscopic properties}
~~~~$\teffstar$ (K)\dotfill         &  \hatcurSMEteff\phn\phn   & SME \tablenotemark{a}\\
~~~~$\feh$ \dotfill                 &  \hatcurSMEzfeh\phs   & SME                 \\
~~~~$\vsini$ (\kms)\dotfill         &  \hatcurSMEvsin   & SME                 \\
~~~~$\vmac$ (\kms)\dotfill          &  \hatcurSMEvmac   & SME                 \\
~~~~$\vmic$ (\kms)\dotfill          &  \hatcurSMEvmic   & SME                 \\
~~~~$\gamma_{\rm RV}$ (\kms)\dotfill         &  \hatcurDSgamma\phn\phs       & DS      \\
\sidehead{Photometric properties}
~~~~$B_T$ (mag)\dotfill             &  10.494 $\pm$ 0.031\phn  & Tycho-2 \\
~~~~$V_T$ (mag)\dotfill             &  10.038 $\pm$ 0.029\phn  & Tycho-2 \\
~~~~$V$ (mag)\dotfill               &   9.980 $\pm$ 0.058  & TASS \\
~~~~$I_C$ (mag)\dotfill             &   9.453 $\pm$ 0.102  & TASS \\
~~~~$J$ (mag)\dotfill               &   9.094 $\pm$ 0.021  & 2MASS \\
~~~~$H$ (mag)\dotfill               &   8.927 $\pm$ 0.020  & 2MASS \\
~~~~$K_s$ (mag)\dotfill             &   8.851 $\pm$ 0.019  & 2MASS \\
~~~~$E(B\!-\!V)$ (mag)\dotfill      &   0.028 $\pm$ 0.015  & Dust maps \tablenotemark{b} \\
\sidehead{Derived properties}
~~~~$\mstar$ ($\msun$)\dotfill      &  \hatcurISOmlong   & \hatcurisoshort+\hatcurlumind+SME \tablenotemark{c}\\
~~~~$\rstar$ ($\rsun$)\dotfill      &  \hatcurISOrlong   & \hatcurisoshort+\hatcurlumind+SME         \\
~~~~$\loggstar$ (cgs)\dotfill       &  \hatcurISOlogg    & \hatcurisoshort+\hatcurlumind+SME         \\
~~~~$\lstar$ ($\lsun$)\dotfill      &  \hatcurISOlum     & \hatcurisoshort+\hatcurlumind+SME         \\
~~~~$M_V$ (mag)\dotfill             &  \hatcurISOmv      & \hatcurisoshort+\hatcurlumind+SME         \\
~~~~Age (Gyr)\dotfill               &  \hatcurISOage     & \hatcurisoshort+\hatcurlumind+SME         \\
~~~~Distance (pc)\dotfill           &  \hatcurXdist\phn  & \hatcurisoshort+\hatcurlumind+SME\\ [-1.5ex]
\enddata
\tablenotetext{a}{
	SME = ``Spectroscopy Made Easy'' package for the analysis of
	high-resolution spectra \citep{valenti:1996}. These parameters
	rely primarily on SME, but have a small dependence also on the iterative
	analysis incorporating the isochrone search and global modeling of
	the data, as described in the text.
}
\tablenotetext{b}{Average of results from \cite{schlegel:1998},
\cite{burstein:1982}, and \cite{hakkila:1997}, with small corrections
for distance.}
\tablenotetext{c}{
	\hatcurisoshort+\hatcurlumind+SME = Based on the 
        \hatcurisoshort\ isochrones \citep{yi:2001},
        \hatcurlumind\ as a luminosity indicator, and the SME results.
}
\end{deluxetable}

Our initial values of \teffstar, \loggstar, and \feh\ were used to
determine auxiliary quantities needed in the global modeling of the
follow-up photometry and radial velocities (specifically, the
limb-darkening coefficients). This modeling, the details of which are
described in the next section, uses a Monte Carlo approach to deliver
the numerical probability distribution of \arstar\ and other fitted
variables. For further details we refer the reader to
\cite{pal:2009b}.  When combining \arstar\ (used as a proxy for
luminosity) with assumed Gaussian distributions for \teffstar\ and
\feh\ based on the SME determinations, a comparison with stellar
evolution models allows the probability distributions of other stellar
properties to be inferred, including \loggstar. Here we use the
stellar evolution calculations from the Yonsei-Yale (YY) series by
\cite{yi:2001}.  The comparison against the model isochrones was
carried out for each of 10,000 Monte Carlo trial sets (see
\refsecl{globmod}). Parameter combinations corresponding to unphysical
locations in the \hbox{H-R} diagram (1\% of the trials) were ignored,
and replaced with another randomly drawn parameter set.  The result
for the surface gravity, $\loggstar = \hatcurISOlogg$, is
significantly different from our initial SME analysis, which is not
surprising in view of the strong correlations among \teffstar, \feh,
and \loggstar\ that are often present in spectroscopic
determinations. Therefore, we carried out a second iteration in which
we adopted this value of \loggstar\ and held it fixed in a new SME
analysis (coupled with a new global modeling of the RV and \lcs),
adjusting only \teffstar, \feh, and \vsini. This gave
$\teffstar = \hatcurSMEiiteff$\,K, 
$\feh = \hatcurSMEiizfeh$, and 
$\vsini = \hatcurSMEiivsin$\,\kms,
in which the conservative uncertainties for the first two have been
increased by a factor of two over their formal values, as before.  A
further iteration did not change \loggstar\ significantly, so we
adopted the values stated above as the final atmospheric properties of
the star.  They are collected in \reftabl{stellar}, together with the
adopted values for the macroturbulent and microturbulent velocities.

We note that the effective temperature also changed rather
significantly from the first SME iteration, by about 300\,K.  An
external check on the accuracy of \teffstar\ may be obtained from
color indices derived from existing brightness measurements for
\hatcur{}. Photometric measurements on a standard system are available
from the Tycho-2, 2MASS, and TASS catalogs \citep{hog:2000,
skrutskie:2006, droege:2006}, and are summarized in the second block
of \reftabl{stellar}. We constructed seven different color indices,
and applied color/temperature calibrations by \cite{ramirez:2005}
(available for six of the color indices), \cite{casagrande:2006} (four
color indices), and \cite{gonzalez:2009} (five color indices). These
relations include terms that depend on metallicity.  Because such
photometric temperatures are very sensitive to reddening, we examined
the dust maps by \cite{schlegel:1998} in the direction to \hatcur{}
and obtained $E(B\!-\!V) = 0.030$, after a small correction for
distance using the value of \hatcurXdist\,pc reported below. Two other
reddening estimates of 0.022 and 0.033 were obtained similarly from
\cite{burstein:1982} and \cite{hakkila:1997}, respectively. We adopt
the straight average of the three measures, $E(B\!-\!V) = 0.028 \pm
0.015$, with a conservative uncertainty.

We corrected each of the color indices for reddening, and computed
weighted temperature averages for each set of calibrations, with
weights determined from the individual temperature uncertainties. The
latter include all photometric errors, the uncertainty in both the
reddening and the metallicity, and the scatter of the calibrations
themselves. The seven color indices are not completely independent,
but nevertheless give a useful sense of the scatter. The agreement
between the 15 separate estimates from all three calibrations is quite
satisfactory, the dispersion being 112\,K.  The mean of the three
calibration averages is $T_{\rm eff} = 6598 \pm 100$\,K, which is in
excellent accord with the final spectroscopic value from SME and
supports its accuracy.

\begin{figure}[h]
\epsscale{1.15}
\plotone{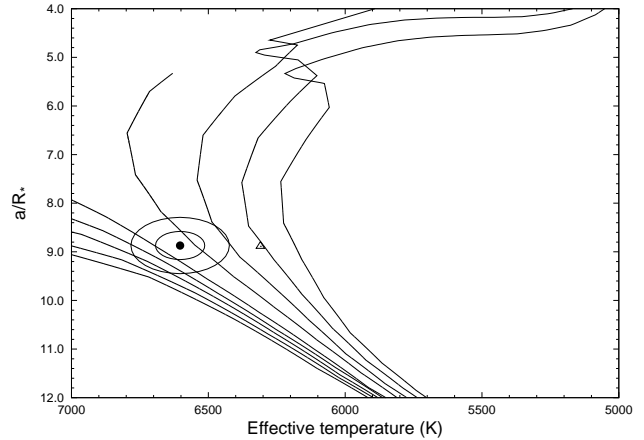}
\caption{
	Model isochrones from \citet{yi:2001} for the measured
        metallicity of \hatcur{}, \feh = \hatcurSMEiizfehshort, and
        ages of 0.2, 0.4, 0.6, 0.8, 1.0, 1.5, 2.0, 2.5, and 3.0\,Gyr
        (left to right).  The adopted values of $\teffstar$ and
        \arstar\ are shown together with their 1$\sigma$ and 2$\sigma$
        confidence ellipsoids. The initial values of \teffstar\ and
        \arstar\ from the first SME and \lc\ analyses are represented
        with a triangle.
\label{fig:iso}}
\end{figure}

With the adopted spectroscopic parameters the model isochrones yield
the stellar mass and radius \mstar\ = \hatcurISOmlong\,\msun\ and
\rstar\ = \hatcurISOrlong\,\rsun, along with other properties listed
at the bottom of \reftabl{stellar}. \hatcur{} is a slightly evolved F
star with an estimated age of \hatcurISOage\,Gyr, according to these
models.  The inferred location of the star in a diagram of \arstar\
versus \teffstar, analogous to the classical H-R diagram, is shown in
\reffigl{iso}. The stellar properties and their 1$\sigma$ and
2$\sigma$ confidence ellipsoids are displayed against the backdrop of
\cite{yi:2001} isochrones for the measured metallicity of \feh\ =
\hatcurSMEiizfehshort, and a range of ages. For comparison, the
location implied by the initial SME results is also shown (triangle),
and corresponds to a somewhat more evolved state.

The distance to the star may be obtained by comparing the absolute
magnitudes inferred from the stellar evolution models against the
apparent brightness measurements mentioned earlier. For this we use
the $V$, $I_C$, $J$, $H$, and $K_s$ magnitudes, corrected for
extinction using $A(V) = 3.1\times E(B\!-\!V)$ appropriately scaled to
each passband following \cite{cardelli:1989}. The near-infrared
magnitudes from 2MASS were transformed to the ESO system of the
isochrones using the relations by \cite{carpenter:2001}. The five
distance estimates agree with each other to within a few pc. The
average is \hatcurXdist\,pc, where the uncertainty includes all
sources of error (photometric and spectroscopic), but excludes
possible systematics from the evolution models themselves, which are
difficult to quantify.

\subsection{Global modeling of the data}
\label{sec:globmod}

This section describes the procedure we followed to model the HATNet
photometry, the follow-up photometry, and the radial velocities
simultaneously.  Our model for the follow-up \lcs\ used analytic
formulae from \citet{mandel:2002} for the eclipse of a star by a
planet, with limb darkening being prescribed by a quadratic law.  The
limb darkening coefficients for the Sloan \band{i} were interpolated
from the tables by \citet{claret:2004} for the spectroscopic
parameters of the star as determined from the SME analysis
(\refsecl{stelparam}). The transit shape was parametrized by the
normalized planetary radius $p\equiv \rpl/\rstar$, the square of the
impact parameter $b^2$, and the reciprocal of the half duration of the
transit $\zrstar$. We chose these parameters because of their simple
geometric meanings and the fact that they show negligible correlations
\citep[see][]{bakos:2009}. The relation between $\zrstar$ and the
quantity \arstar, used in \refsecl{stelparam}, is given by
\begin{equation}
\arstar = P/2\pi (\zrstar) \sqrt{1-b^2} \sqrt{1-e^2}/(1+e \sin\omega)
\end{equation}
\citep[see, e.g.,][]{tingley:2005}.  Our model for the HATNet data was
the simplified ``P1P3'' version of the \citet{mandel:2002} analytic
functions (an expansion in terms Legendre polynomials), for the
reasons described by \citet{bakos:2009}.
Following the formalism presented by \citet{pal:2009}, the RVs were
fitted with an eccentric Keplerian model parametrized by the
semiamplitude $K$ and Lagrangian elements $k \equiv e \cos\omega$ and
$h \equiv e \sin\omega$, in which $\omega$ is the longitude of
periastron.

We assumed that there is a strict periodicity in the individual
transit times. We assigned the transit number $N_{tr} = 0$ to the
first complete follow-up \lc\ gathered on 2009 March 25.  The
adjustable parameters in the fit that determine the ephemeris were
chosen to be the time of the first transit center in HATNet field
G195, $T_{c,-473}$, and that of the last transit center observed with
the \flwof\ telescope, $T_{c,+11}$. We used these as opposed to the
period and reference epoch in order to minimize correlations between
parameters \citep[see][]{pal:2008}. Times of mid-transit for
intermediate events were interpolated using these two epochs and the
corresponding transit number of each event, $N_{tr}$.  The eight main
parameters describing the physical model were thus $T_{c,-473}$,
$T_{c,+11}$, $\rpl/\rstar$, $b^2$, $\zrstar$, $K$, $k \equiv
e\cos\omega$, and $h \equiv e\sin\omega$. Five additional ones were
included that have to do with the instrumental configuration. These
are the HATNet blend factors $B_{\rm inst}$ (one for each HATNet
field, G194 and G195), which account for possible dilution of the
transit in the HATNet \lcs\ from background stars due to the broad PSF
(20\arcsec\ FWHM), the HATNet out-of-transit magnitudes $M_{\rm
0,HATNet}$ (one for each field), and the relative zero-point
$\gamma_{\rm rel}$ of the Keck RVs.

We extended our physical model with an instrumental model that
describes brightness variations caused by systematic errors in the
measurements. This was done in a similar fashion to the analysis
presented by \citet{bakos:2009}. The HATNet photometry had already
been EPD- and TFA-corrected before the global modeling, so we only
considered corrections for systematics in the follow-up \lcs. We chose
the ``ELTG'' method, i.e., EPD was performed in ``local'' mode with
EPD coefficients defined for each night, and TFA was performed in
``global'' mode using the same set of stars and TFA coefficients for
all nights.  The five EPD parameters were the hour angle (representing
a monotonic trend that changes linearly over time), the square of the
hour angle (reflecting elevation), and the stellar profile parameters
(equivalent to the FWHM, elongation, and position angle of the image).
The functional forms of the above effects contained six coefficients,
including the auxiliary out-of-transit magnitude of the individual
events. The EPD parameters were independent for all five nights,
implying 30 additional coefficients in the global fit. For the global
TFA analysis we chose 20 template stars that had good quality
measurements for all nights and on all frames, implying an additional
20 parameters in the fit. Thus, the total number of fitted parameters
was 13 (physical model with 5 configuration-related parameters) + 30
(local EPD) + 20 (global TFA) = 63, i.e., much smaller than the total
number of data points (1711).

The joint fit was performed as described in \citet{bakos:2009}.  We
minimized \chisq\ in the space of parameters by using a hybrid
algorithm, combining the downhill simplex method \citep[AMOEBA;
see][]{press:1992} with a classical least squares algorithm.
Uncertainties for the parameters were derived applying the Markov
Chain Monte-Carlo method \citep[MCMC, see][]{ford:2006} using
``Hyperplane-CLLS'' chains \citep{bakos:2009}. This provided the full
{\em a posteriori} probability distributions of all adjusted
variables. The length of the chains was 10,000 points. The {\em a
priori} distributions of the parameters for these chains were chosen
to be Gaussian, with eigenvalues and eigenvectors derived from the
Fisher covariance matrix for the best-fit solution. The Fisher
covariance matrix was calculated analytically using the partial
derivatives given by \citet{pal:2009}.

\section{Results}
\label{sec:results}
With the procedure above we obtained the {\em a posteriori}
distributions for all fitted variables, and other quantities of
interest such as \arstar. As described in \refsecl{stelparam},
\arstar\ was used together with stellar evolution models to infer a
theoretical value for \loggstar\ that is significantly more accurate
than the spectroscopic value. The improved estimate was in turn
applied to a second iteration of the SME analysis, as explained
previously, in order to obtain better estimates of \teffstar\ and
\feh.  The global modeling was then repeated with updated
limb-darkening coefficients based on those new spectroscopic
determinations. The resulting geometric parameters pertaining to the
light curves and velocity curves are listed in
\reftabl{planetparam}. The rms residual of the RV measurements from
the orbital fit is 8.1\,\ms.

\begin{deluxetable}{lc}
%\tablewidth{0pc}
\tabletypesize{\scriptsize}
\tablecaption{Orbital and planetary parameters\label{tab:planetparam}}
\tablehead{
	\colhead{~~~~~~~~~~~Parameter~~~~~~~~~~~} &
	\colhead{Value}
}
\startdata
\noalign{\vskip -3pt}
\sidehead{\Lc{} parameters}
~~~$P$ (days)             \dotfill    & $\hatcurLCP$              \\
~~~$T_c$ (${\rm BJD}$)
      \tablenotemark{a}    \dotfill    & $\hatcurLCT$              \\
~~~$T_{14}$ (days)
      \tablenotemark{a}   \dotfill    & $\hatcurLCdur$            \\
~~~$T_{12} = T_{34}$ (days)
    \tablenotemark{a}     \dotfill    & $\hatcurLCingdur$         \\
~~~$\arstar$              \dotfill    & $\hatcurPPar$             \\
~~~$\zrstar$              \dotfill    & $\hatcurLCzeta$\phn           \\
~~~$\rpl/\rstar$          \dotfill    & $\hatcurLCrprstar$        \\
~~~$b^2$                  \dotfill    & $\hatcurLCbsq$            \\
~~~$b \equiv a \cos i/\rstar$
                          \dotfill    & $\hatcurLCimp$            \\
~~~$i$ (deg)              \dotfill    & $\hatcurPPi$\phn         \\
\sidehead{Limb-darkening coefficients (Sloan \band{i}) \tablenotemark{b}}
~~~~$c_1$ (linear term)\dotfill			&  \hatcurLBii	\\%	& SME+Claret\tablenotemark{b}  \\
~~~~$c_2$ (quadratic term)\dotfill			&  \hatcurLBiii	\\%	& SME+Claret          \\

\sidehead{RV parameters}
~~~$K$ (\ms)              \dotfill    & $\hatcurRVK$\phn\phn              \\
~~~$k \equiv e \cos\omega$ \tablenotemark{c} 
                          \dotfill    & $\hatcurRVk$\phs              \\
~~~$h \equiv e \sin\omega$ \tablenotemark{c}
                          \dotfill    & $\hatcurRVh$              \\
~~~$e$                    \dotfill    & $\hatcurRVeccen$          \\
~~~$\omega$ (deg)            \dotfill    & $\hatcurRVomega$\phn   \\
~~~RV jitter (\ms)           \dotfill    & \hatcurRVjitter           \\
~~~$\sigma_{\rm RV}$ (\ms)           \dotfill    & 8.1           \\

\sidehead{Planetary parameters}
~~~$\mpl$ ($\mjup$)       \dotfill    & $\hatcurPPmlong$          \\
~~~$\rpl$ ($\rjup$)       \dotfill    & $\hatcurPPrlong$          \\
~~~$C(\mpl,\rpl)$
    \tablenotemark{d}     \dotfill    & $\hatcurPPmrcorr$         \\
~~~$\rhopl$ (\gcmc)       \dotfill    & $\hatcurPPrho$            \\
~~~$\log g_p$ (cgs)       \dotfill    & $\hatcurPPlogg$           \\
~~~$a$ (AU)               \dotfill    & $\hatcurPParel$           \\
~~~$T_{\rm eq}$ (K)       \dotfill    & $\hatcurPPteff$\phn\phn           \\
~~~$\Theta$\tablenotemark{e}               \dotfill    & $\hatcurPPtheta$          \\
~~~$\langle F \rangle$ ($10^9$ \ergscmsq) \tablenotemark{f}
                          \dotfill    & $\hatcurPPfluxavg$        \\ [-1.5ex]
\enddata
\tablenotetext{a}{
        \ensuremath{T_c}: Reference epoch of mid transit that minimizes the
        correlation with the orbital period. It corresponds to $N_{tr} = -9$.
	\ensuremath{T_{14}}: total transit duration, i.e., the interval of time
	between the first and last contacts;
	\ensuremath{T_{12}=T_{34}}: ingress/egress time, i.e., the time interval
	between the first and second, or third and fourth contacts.
}
\tablenotetext{b}{Values for a quadratic law, adopted from the tabulations
by \cite{claret:2004} according to the spectroscopic (SME) parameters listed in
\reftabl{stellar}.}
\tablenotetext{c}{
    Lagrangian orbital parameters derived from the global modeling, 
    and primarily determined by the RV data. 
}
\tablenotetext{d}{
	Correlation coefficient between the planetary mass \mpl\ and radius
	\rpl.
}
\tablenotetext{e}{
	Safronov number given by $\Theta = \frac{1}{2}(V_{\rm
	esc}/V_{\rm orb})^2 = (a/\rpl)(\mpl / \mstar )$
	\citep[see][]{hansen:2007}.
}
\tablenotetext{f}{
	Incoming flux per unit surface area, averaged over the slightly
eccentric orbit.
}
\end{deluxetable}

Included in this table is the RV ``jitter''. This is a component of
assumed astrophysical noise intrinsic to the star that we added in
quadrature to the internal errors for the RVs in order to achieve
$\chi^{2}/{\rm dof} = 1$ from the RV data for the global
fit. Auxiliary parameters not listed in the table are:
$T_{\mathrm{c},-473}=\hatcurLCTA$~(BJD),
$T_{\mathrm{c},+11}=\hatcurLCTB$~(BJD), the blending factors
$B_{\rm instr}=\hatcurLCiblendnew$ for HATNet field G194 and
$B_{\rm instr}=\hatcurLCiblendold$ for field G195, and
$\gamma_{\rm rel}=\hatcurRVgamma$\,\ms. 
The latter quantity represents an arbitrary offset for the Keck RVs,
and does \emph{not} correspond to the true center of mass velocity of
the system, which was listed earlier in \reftabl{stellar}
($\gamma_{\rm RV}$).

The planetary parameters and their uncertainties can be derived by
combining the {\em a posteriori} distributions for the stellar, \lc,
and RV parameters.  In this way we find a mass for the planet of
$\mpl=\hatcurPPmlong\,\mjup$ and a radius of
$\rpl=\hatcurPPrlong\,\rjup$, leading to a mean density
$\rho_p=\hatcurPPrho$\,\gcmc. These and other planetary parameters are
listed at the bottom of Table~\ref{tab:planetparam}.

We note that the eccentricity of the orbit is significantly different
from zero: $e = \hatcurRVeccen$. With a longitude of periastron
$\omega = 95\arcdeg \pm 4\arcdeg$, the orientation of the system is
such that the orbit is viewed very nearly along the major axis, with
periastron on the near side. Furthermore, the eccentricity is large
enough that the separation between the star and the planet when the
planet is behind is some 23\% larger than at transit. This, combined
with the relatively low inclination angle, implies that there should
be no occultations of \hatcur{b}. Formally, the impact parameter at
the occultation is $b_{\rm occult} = 1.102 \pm 0.030$, which is
greater than unity at the 3$\sigma$ confidence level.

Individual times of mid transit were measured for each of the five
events observed with KeplerCam. This was done by holding fixed all
model parameters that define the shape of the transit to their final
values in \reftabl{planetparam}, and adjusting only the time of the
center of the event in each series. The results are listed in
\reftabl{timings}, along with their uncertainties and $O\!-\!C$
residuals from the ephemeris for the planet. As expected, the events
with only partial coverage of the transit (first two in
\reftabl{timings}; see also \reffigl{lc}) have larger uncertainties
and residuals.

\begin{deluxetable}{lccc}
\tablewidth{0pc}
\tablecaption{Transit timing measurements for \hatcur\label{tab:timings}}
\tablehead{
        \colhead{~~~~~~~~$N_{tr}$~~~~~~~~} &
	\colhead{$T_{\rm c}$} & 
	\colhead{$\sigma_{T_{\rm c}}$} & 
	\colhead{$O\!-\!C$} \\
        \colhead{} &
	\colhead{\hbox{(BJD$-$2,454,000)}} & 
	\colhead{(days)} & 
	\colhead{(sec)}
}
\startdata
 $-34$\dotfill  &  759.59280  & 0.00290 & $-$458     \\
 $-11$\dotfill  &  866.03268  & 0.00273 & $-$127     \\
 \phn+0\dotfill &  916.93940  & 0.00081 & \phn$+$90      \\
 \phn+3\dotfill &  930.82160  & 0.00114 & \phn$+$24      \\
 +11\dotfill    &  967.84186  & 0.00088 & \phn$-$59      \\ [-1.5ex]
\enddata
\end{deluxetable}

As a final note, the barycentric Julian dates in this table and
throughout the paper are calculated from Coordinated Universal Time
(UTC), in which a second has the same duration as a second of
International Atomic Time (TAI), but the scale is subject to
occasional 1-second adjustments (leap seconds) to stay in pace with
the slowly and irregularly decreasing rotation rate of the Earth. UTC
(the basis of civil time) is therefore not continuous, and is
unsuitable as a physical time coordinate.  This can be important for
some applications that combine timing measurements over long periods
of time \citep[e.g.,][]{Guinan:2001}, a problem that has long been
recognized in the binary star community \citep[for a pointed
discussion, see][]{Bastian:2000} as well as the Solar System and
pulsar communities. Timing measurements in the transiting planet field
now span a decade, and although the differential effect over this
period is only a few seconds (significantly smaller than the typical
measurement precision), it will become more important in the future.
The issue has already been mentioned recently in the context of
studying transit timing variations for transiting planets
\citep{Adams:2010}.  While the use of Terrestrial Time (TT = TAI +
32.184 seconds) as the basis for Julian date calculations eliminates
the problem, for practical reasons virtually all the JDs (and BJDs)
reported in previous transiting planet studies, including those in the
HATNet series, are implicitly on the UTC scale. We continue this
practice here, but caution the reader that the TT$-$UTC corrections
need to be taken into account for the most demanding applications.

\section{Discussion}
\label{sec:discussion}

At $V = \hatcurCCtassmv$, \hatcur{} is among the brightest TEP host
stars presently known (only 11 of the other 66 are brighter), which
should facilitate a wide range of follow-up studies.  Our planetary
mass determination of $\mpl\ = \hatcurPPmshort$\,\mjup\ places
\hatcur{b} in the category of the massive transiting planets
discovered to date, which seem to be less common than those with
smaller masses.  In fact, a look at the mass distribution for the
known transiting systems shows that it falls off precisely around
2\,\mjup. Since there is no obvious selection effect that would make
these massive objects more difficult to detect or confirm (such as
smaller radii, longer orbital periods, or fainter or hotter parent
stars), it may be that they are indeed less common as a population.
It has also been noticed \citep[e.g.,][]{southworth:2009, joshi:2009}
that the most eccentric cases among the transiting planets are all
massive: all systems with $e > 0.2$ are more massive than 3\,\mjup,
although the numbers are still small.  A similar trend is seen among
the non-transiting planets, bearing in mind the $\sin i$ ambiguity in
their masses. Among the transiting systems whose spin-orbit alignment
has been measured through observation of the Rossiter-McLaughlin
effect, it would also appear that the more massive ones are
disproportionately misaligned compared to the ones with smaller-mass
planets, considering their relative populations, although again the
numbers are as yet too small to be conclusive. Obviously it would be
useful to determine the spin-orbit alignment of the \hatcur{} system,
which should be readily measurable given that $v \sin i = 8.4$\,\kms.

The mass of \hatcur{b} is quite similar to the recently discovered
Kepler-5b \citep{koch:2010}, but its radius is about 20\% smaller. A
comparison with the evolution models of \cite{fortney:2007}, at an
equivalent solar semimajor axis of $a_{\rm equiv} = 0.0317$\,AU,
indicates that the radius of \hatcur{b} is well reproduced for the
\hatcurISOageshort\,Gyr age of the system if it has about 50\,\mearth\
worth of heavy elements in its interior (about 7\% of its total
mass). This amount of metals is consistent with the correlation
between core mass and metallicity of the parent star proposed by
\cite{Guillot:2006} and \cite{Burrows:2007}, which seems to support
the core-accretion mode of planet formation.  The incident flux we
compute for the planet averaged over its eccentric orbit, $\langle F
\rangle = \hatcurPPfluxavg \times 10^9$ \ergscmsq, places it in the
proposed pM category of \cite{fortney:2008}. These objects are
expected to present temperature inversions in their atmospheres, and
large day/night temperature contrasts. The equilibrium temperature of
\hatcur{b}, assuming zero albedo and full redistribution of the
incident radiation, is \hatcurPPteff\,K.

The non-zero eccentricity of the orbit for the relatively short period
of 4.6 days raises the question of whether tidal forces have had
enough time to circularize the orbit at the \hatcurISOage\,Gyr age of
the system, or alternatively, whether there might be a second planet
perturbing the orbit of \hatcur{b} and pumping up its eccentricity.
Following \cite{adams:2006}, we estimate the timescale for tidal
circularization to be $\tau_{\rm circ} \approx 0.33$\,Gyr, although
this is a strong function of the poorly known tidal quality factor
$Q_p$, for which we have adopted here the commonly used value of
$10^5$. For $Q_p \simeq 10^6$ the timescale is 10 times longer, and a
recent study of the 0.78-day transiting planet WASP-19b has suggested
$Q_p$ could be much larger still \citep{hebb:2010}.  Consequently we
cannot rule out that the eccentricity we measure for \hatcur{b} is
primordial, at least in part, nor can we exclude the presence of a
second planet in the system.

The grazing orientation ($b = \hatcurLCimp$) significantly enhances
the detection sensitivity to additional planets in the system. These
could manifest themselves through transit {\it duration} variations
(TDVs), or equivalently, changes in the orbital inclination angle as
has been claimed for TrES-2b by \cite{mislis:2009a} and
\cite{mislis:2009b}, and recently disputed by \cite{scuderi:2010}.
Perturbing planets may also induce variations in the times of
mid-transit (transit {\it timing} variations, or TTVs) through
gravitational interaction \citep[e.g.,][]{Dobrovolskis:1996,
Holman:2005}. In general, there exists no exact closed-form
expressions for evaluating the amplitudes of such signals since no
general solution to the 3-body problem exists. One typical approach is
to employ an approximation to simplify the problem. \cite{Agol:2005}
provided useful analytic expressions for the case of a coplanar system
where the host star has a much larger mass than the two companion
planets, for several different scenarios.  The largest TTVs occur for
planets in mean motion resonance, and we may use expressions [33] and
[34] from \cite{Agol:2005} to estimate the amplitude and libration
period of the TTV signal in a 2:1 resonance.  For the \hatcur{} system
we find that such a planet would induce a peak-to-peak deviation in
the timings of 13.6\,s and 126.3\,s for a mass of $0.1\,\mearth$ and
$1\,\mearth$, respectively, and the libration period would be
$\sim$175 days. Thus, an Earth-mass planet in this configuration would
be eminently detectable with further observations.  Out-of-resonance
perturbers can be considered using equation [32] of \cite{Agol:2005},
but we find they lead to sub-second amplitudes for various
configurations, far beyond the current measurement capabilities.

The properties of \hatcur{b} also make it attractive for a future
search for a companion satellite (`exomoon').  The Hill radius ($R_H$)
extends to 8.1 planetary radii (9.1\,$R_J$), and using the expressions
of \citet{domingos:2006}, which account for the planet's orbital
eccentricity, the maximum distance at which an exomoon could be stable
for this planet would be 0.436\,$R_H$ and 0.824\,$R_H$ for a prograde
and retrograde exomoon, respectively.  \citet{barnes:2002} presented
analytic approximations for the maximum stable exomoon mass for
close-in extrasolar giant planets based upon tidal dissipation
arguments, which were shown to provide excellent agreement with
numerical integrations.  Assuming again a tidal dissipation value of
$Q_p \simeq 10^5$ and a Love number of $k_{2p} \simeq 0.51$, the
maximum stable moon mass over the \hatcurISOageshort\,Gyr lifetime of
the star is 0.002\,\mearth\ and 0.12\,\mearth\ for a moon at the
maximum planet-moon separations for prograde and retrograde orbits,
respectively. For $Q_p \simeq 10^6$ these limits would be 10 times
larger, or 1.2\,\mearth\ in the retrograde case. Such exomoons would
produce small TTVs, but the highly grazing orientation provides a
better means of detection.  \citet{kipping:2009a, kipping:2009b}
showed that an exomoon induces two types of transit duration
variations on the host planet: a sky-projected tangential velocity
variation (the V-component), and a spatially orthogonal transit impact
parameter variation (the TIP component). While the TTV and TDV-V rms
amplitudes of a 1.2\,\mearth\ exomoon would be 4.5\,s and 1.0\,s,
respectively, the TDV-TIP component could be as high as 16\,s or 45\,s
peak-to-peak for the optimum configuration of a favorably inclined
exomoon.  For comparison, another near-grazing transiting planet is
TrES-2b \citep{odonovan:2006}, and repeating the same calculation
(with $Q_p \simeq 10^6$, as above) we find that the maximal TDV-TIP of
the maximum-mass stable retrograde exomoon would be only 0.07\,s. The
much smaller value in this case is mostly due to the low maximum
stable moon mass of 0.005\,\mearth, as a result of the lower planetary
mass and longer system age.  We suggest, therefore, that future
monitoring of the transit duration of \hatcur{b} could allow for
interesting constraints not only on the presence of additional
planets, but also of sub-Earth mass or Earth mass exomoons.

\acknowledgements 

HATNet operations have been funded by NASA grants NNG04GN74G and
NNX08AF23G, and SAO IR\&D grants. G.T.\ acknowledges partial support
from NASA grant NNX09AF59G. G.\'A.B.\ and J.A.J.  were supported by
Postdoctoral Fellowships of the NSF Astronomy and Astrophysics Program
(AST-0702843 and AST-0702821, respectively). We acknowledge partial
support also from the Kepler Mission under NASA Cooperative Agreement
NCC2-1390 (D.W.L., PI). G.K.\ thanks the Hungarian Scientific Research
Foundation (OTKA) for support through grant K-81373.  G.\'A.B.~wishes
to thank G\'abor Kov\'acs for his help in system management of the
HATNet computers while the data analysis was carried out. We are
grateful to the anonymous referee for helpful suggestions.  This
research has benefited from Keck telescope time allocations granted
through NOAO (programs A264Hr, A146Hr) and NASA (N049Hr, N018Hr).
This research has also made use of the SIMBAD database and the VizieR
catalogue access tool, both operated at CDS, Strasbourg, France, of
NASA's Astrophysics Data System Abstract Service, and of data products
from the Two Micron All Sky Survey (2MASS), which is a joint project
of the University of Massachusetts and the Infrared Processing and
Analysis Center/California Institute of Technology, funded by NASA and
the NSF.

\end{document}